\documentclass[twocolumn,preprintnumbers,amsmath,amssymb, floatfix]{revtex4}

\usepackage{graphicx}
\usepackage{dcolumn}
\usepackage{bm}

\makeatletter
\def\@dotsep{4.5}
\makeatother

\begin{document}

\title{Asymmetric parametric amplification in nonlinear left-handed transmission lines}

\author{David A. Powell}
\email{david.a.powell@anu.edu.au}
\author{Ilya V. Shadrivov}
\author{Yuri S. Kivshar}
\affiliation{Nonlinear Physics Centre, Research School of Physics and Engineering, Australian National University, Canberra ACT 0200, Australia}%

\begin{abstract}
We study parametric amplification in nonlinear left-handed transmission lines, which serve as model systems for nonlinear negative index metamaterials.  We experimentally demonstrate amplification of a weak pump signal in three regimes: with the signal in the left-handed band, with the signal in the stop band, and with the signal at a defect frequency.  In particular, we demonstrate  the amplification of the incident wave by up to 15dB in the left-handed regime.
\end{abstract}

\maketitle

Left-handed transmission lines are compact systems showing regimes of backward-wave propagation similar to negative-index metamaterials.  They have been applied to a number of engineering applications, including the study of leaky wave antennas, compact resonators, and dual-band couplers (see, e.g., Ref.~\cite{Lai2004} and references therein).  In a number of these applications nonlinear elements have been introduced to create tunable structures~\cite{Gil2006,Kuylenstierna2006} and, in addition, they have been used as a platform for the study of nonlinear wave propagation in the systems supporting the propagation of backward waves~\cite{Kozyrev2005,Kozyrev2007a,Kozyrev2008}.

Parametric gain is a nonlinear process whereby a high-energy pump wave exchanges energy with a weaker signal wave through modulation of the material or circuit parameters, thus amplifying it.  This effect is commonly used in optical systems, and it has been proposed as a way to mitigate the losses in negative-index metamaterials.  In particular, in Ref.~\cite{Kozyrev2006} parametric gain was exhibited in the left-handed region for a short structure consisting of seven periods with an external DC bias, and parametric generation of backward waves in similar transmission lines was studied both experimentally and theoretically in Ref.~\cite{Gorshkov1998}.

Recently, we have demonstrated a bistable regime of left-handed propagation in an asymmetric nonlinear left-handed transmission line due to the existence of multiple dynamic states~\cite{Powell2008a}. In this Letter, we demonstrate that such structures may exhibit substantial parametric gain in three different regimes of the signal wave: with the signal in the left-handed band, with the signal in the stop band, and with the signal at a defect frequency.  In particular, we demonstrate  the amplification of the incident wave by up to 15dB in the left-handed regime.

\begin{figure}[htb]
\centering
\includegraphics[width=0.95\columnwidth]{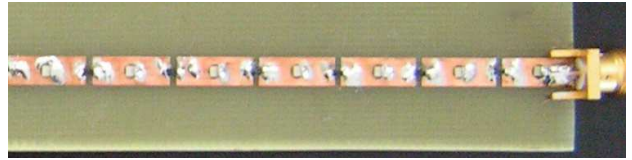}
\caption{\label{fig:photo}Photograph showing six periods of the nonlinear left-handed transmission line
used in our experiments.}
\end{figure}

\begin{figure}[htb]
\centering
\includegraphics[width=\columnwidth]{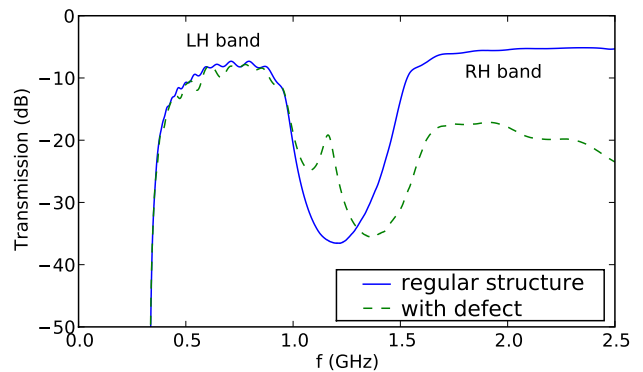}
\caption{\label{fig:transmission}Transmission response of the left-handed transmission line, before and after the inclusion of a defect.}
\end{figure}

Our nonlinear transmission line consists of 20 periods of elementary cells, each cell containing a series varactor diode (SMV1405), shunt inductor (18nH chip type), and a 10~mm section of 50$\mathrm{\Omega}$ microstrip transmission line, as shown in Fig. \ref{fig:photo}.  The resultant transmission response is presented by a solid curve in Fig.~\ref{fig:transmission}. The dispersion relation is of great importance to all parametric processes, and it can be calculated analytically as \cite{Lai2004}:
\begin{eqnarray}\label{eq:dispersion}
\cos\kappa = \left[1 - \frac{1}{2}\left(\omega L_{R} - \frac{1}{\omega C_{L}}\right)\left(\omega C_{R} - \frac{1}{\omega L_{L}}\right)\right],
\end{eqnarray}
where $\kappa$ is the Bloch wavenumber (in radians per period), $L_{L}$ and $C_{L}$ are the shunt inductance and series capacitance introduced to create the left-handed line; $C_{R} = d/Z_{0}v$ and $L_{R} = Z_{0}d/v$ are the equivalent shunt capacitance and series inductance of the microstrip line having impedance $Z_{0}$ and propagation velocity $v$, calculated under the assumption that each section length $d$ is short compared to the wavelength.  Such structures demonstrate a transmission spectrum with a low-frequency stop band, a left-handed pass band, where the phase velocity and Poynting vector are in opposite directions, another stop band (which may be absent if the left- and right-handed bands edges coincide), and a right-handed pass band where the phase velocity and Poynting vector are co-directional.

\begin{figure}[htb]
\centering
\includegraphics[width=\columnwidth]{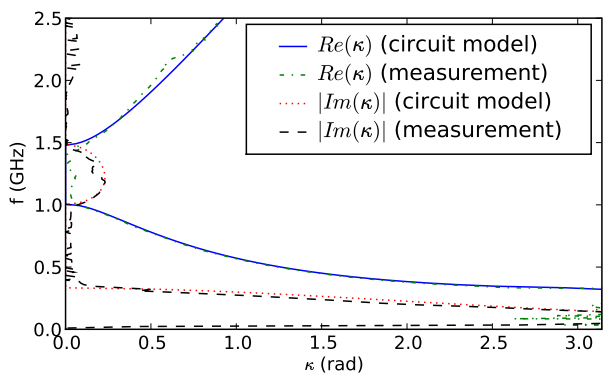}
\caption{\label{fig:dispersion}Dispersion curves fitted to experimental data using the retrieval procedure in Ref.~\cite{Sukhorukov2008} and from the circuit theory.}
\end{figure}

We first verify the dispersion of the transmission line experimentally. One of the standard approaches is to scan the fields above the structure for each frequency and perform a Fourier transform in the spatial domain in order to find the dominant spatial frequencies.  However, for a structure such as ours which is only 20 periods long, this approach gives a very poor resolution of spatial frequencies.  Instead, we make the \emph{a priori} assumption that only a single Bloch mode propagates in the structure at any given frequency.  We then perform an optimization procedure to find its wavenumber, as well as the amplitudes of the forward and backward propagating waves, using the procedure described in Ref.~\cite{Sukhorukov2008}.  Importantly, our procedure allows retrieval of both the real and imaginary parts of the wavenumber. The latter characterizes the decay length of the waves inside and outside of the band-gaps.  The corresponding results are shown in Fig.~\ref{fig:dispersion}, and they are compared with Eq.~\eqref{eq:dispersion}, with the parasitic inductance of the diode added to the term $L_R$.

It can be seen that agreement is quite good over the whole frequency range, and that the locations of the band edges are also in agreement with the theory.  Some additional imaginary parts of the wavenumber in the transmission bands are seen in the experiment due to losses which are not included in the simplified circuit model.  The relative amplitudes of the forward and backward Bloch modes (not shown) verify the regions of the forward and backward-wave propagation.  For high input powers we do not observe any smooth shift of this dispersion curve, instead artifacts appear as the wave no longer satisfies the Bloch condition, and the dispersion curve becomes meaningless.

In order to achieve strong parametric amplification, it is necessary to ensure that the generated signal adds to itself constructively over the length of the structure.  In three-frequency parametric processes with all components forward-propagating, this will occur when $\omega_{1}+\omega_{2} = \omega_{3}$ and $\kappa_{1}+\kappa_{2} = \kappa_{3}$.  For any frequency component where the backward wave is dominant a negative sign should appear in front of $\kappa$.  Here subscript 1 indicates the signal wave, subscript 3 indicates the pump wave, and subscript 2 indicates the idler wave, which is automatically generated through the parametric process.  However, it should be noted that for relatively short structures with strong nonlinear interaction and reflection from the structure edges such as our transmission line, parametric amplification can occur which satisfies different conditions for the wavenumbers~\cite{Shen:1984:PrinciplesNonlinear}. These conditions may depend on the length of the structure and reflections from the boundaries.

In order to measure parametric gain, the signal and pump are combined with a 3dB combiner.  Since the combiner does not offer good isolation over all frequencies, a 10dB attenuator is applied to the signal line to provide isolation of the source generator from the pump signal.  The pump and signal frequencies are determined empirically to find the frequencies of maximum gain.  The pump is applied at $f_3=1.43$GHz, which is inside the stop-band and thus has $\kappa_3=0$.  The signal is applied in the left-handed band at $f_1 = 806$MHz with $\kappa_1=-0.45$rad, and it is co-directional to the excited idler with $f_2 = 624$MHz and $\kappa_2=-0.85$rad. 

\begin{figure}[htb]
\centering
\includegraphics[width=\columnwidth]{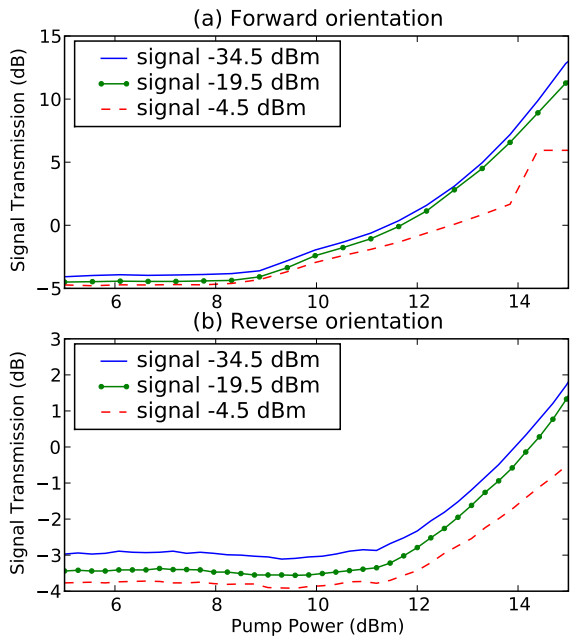}
\caption{\label{fig:lh}Gain vs. pump power for the signal in the left-handed band, for two different orientations of the transmission line.}
\end{figure}

Gain is measured by taking the ratio of the signal power at the output of the transmission line to the signal power incident at the input, thus it also includes the effects of the losses in the line.  Figure~\ref{fig:lh}(a) shows the gain of the transmission line as a function of the pump power incident at the input port of the transmission line.  At low powers the pump has essentially no effect, and the attenuation of 4-5dB due to the loss through the transmission line is observed.  Once the pump power reaches he value ~12dBm it can be seen that the pump compensates for the losses of the line, and subsequently the system provides gain, up to 15dB for pump powers of 15dBm.  Beyond this point the system becomes a parametric generator.  Also shown is the influence of the signal power, which reduces the gain and for very high input powers the gain process is not so clearly observable due to interaction with the many intermodulation products present in the system.

In this structure, there is only a single diode per input cell, and they are all identically aligned, thus there is a strong asymmetry in the system response.  This asymmetry strongly affects the parametric gain, and when both the signal and pump are applied in the other port of the transmission line, the parametric gain is much reduced, as shown in Fig.~\ref{fig:lh}(b).  This may be due to the onset of forward conduction, or the asymmetry of the capacitance-voltage relationship of the diodes.

\begin{figure}[htb]
\centering
\includegraphics[width=\columnwidth]{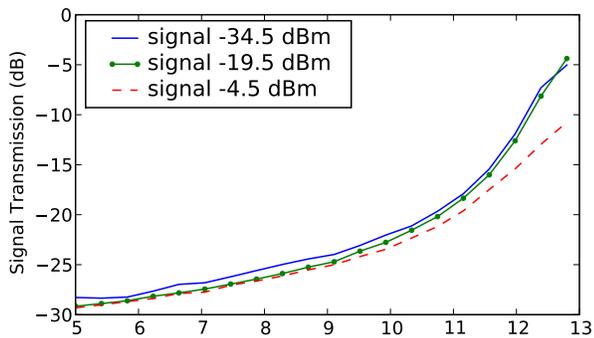}
\caption{\label{fig:bandgap}Gain vs. pump power for the signal within the band gap.}
\end{figure}

We also performed a thorough investigation of possible regimes of parametric gain.  We were not able to find any parametric processes with the signal operating in the right-handed band.  This is not unexpected, since at these frequencies the series capacitance acts as a very small impedance, and has little ability to modulate the propagating wave.  However, it was found that a signal \emph{within the stop-band} could be amplified by a pump in the right-handed band.  This is shown in Fig. \ref{fig:bandgap}, where the pump was at $f_3=1.958$GHz, $k_3 = 0.46$rad, the signal at $f_1=1.376$GHz, $\kappa_1=0$ and idler at $f_2=582$MHz, $\kappa_2=-0.97$rad. 

Finally, we consider the effects associated with the introduction of a defect into the transmission line.  Defects within a periodic system have the ability to localize waves, and they are often used to create resonant cavities at frequencies inside the band gaps.  In our case, the defect is created by introducing an additional inductor in series with the diode.  This creates a defect mode well within the stop band, as well as a degradation of the transmission in the right-handed band, as shown by the dashed curve in Fig.~\ref{fig:transmission}.  A field scan was performed (not shown) to demonstrate that the fields at this frequency are localized on the defect site, and that they decay strongly with distance.

\begin{figure}[htb]
\centering
\includegraphics[width=\columnwidth]{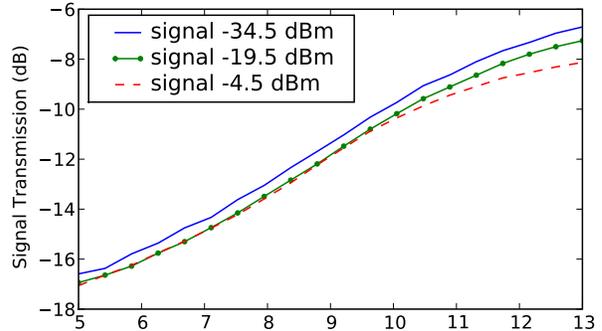}
\caption{\label{fig:defect}Gain vs. pump power for the signal at the frequency of the defect mode.}
\end{figure}

Parametric amplification is performed with a signal at the defect with $f_1=1.164$GHz, $\kappa_1=0$, idler in the left-handed band with $f_2=730$MHz, $\kappa_2=-0.60$rad and pump in the right-handed band at $f_3=1.893$GHz, $\kappa_1=0.41$rad, and the results are summarized in Fig.~\ref{fig:defect}.  It can be seen that the gain is much lower and never becomes positive, which can be attributed to the defect cell strongly reflecting the signal in the right-handed band, thus inhibiting transmission of the pump. 

In conclusion, we have experimentally studied parametric amplification and generation in nonlinear left-handed transmission lines. By utilizing the complex dispersion characteristics of this system, we have demonstrated amplification of a weak pump signal in the three different regimes: with the signal in the left-handed propagation band, with the signal in the stop band, and with the signal at a defect frequency. By engineering the dispersion of such structures, it should be possible to further enhance the gain which can be produced, for example by shifting the pump frequency to the band edge, where the real wavenumber is still zero, but the transmission is much higher. Our results show that substantial gain can be achieved in nonlinear systems exhibiting backward-wave propagation, and they confirm that parametric processes are promising candidates for mitigating the losses found in artificial backward wave media, including negative-index metamaterials.

The authors acknowledge the financial support of the Australian Research Council.



\end{document}